# Mesoporous Silica Nanoparticles-based Smart Nanocarriers for Targeted Drug Delivery in Colorectal Cancer Therapy

Rochelle A. Mann[a], Md. Emran Hossen[b], Alexander David McGuire Withrow[c], Jack Thomas Burton[d], Sean M Blythe[d] and Camryn Grace Evett[e]

[a]Department of Chemistry and Biochemistry, University of California Santa Barbara, Santa Barbara, CA, USA
[b]Department of Biomedical Engineering, Gannon University, Erie, PA, USA
[c]Department of Chemistry, Biochemistry and Physics, South Dakota State University, Brookings, SD, USA
[d]Department of Chemistry, Western Washington University, Bellingham, WA, USA
[e]Department of Medical Biology, University of South Dakota, Vermillion, SD, USA

Corresponding Author: Camryn Grace Evett; E-mail: camryn.evett@coyotes.usd.edu

**ABSTRACT**: Colorectal cancer (CRC) remains a leading cause of cancer-related mortality worldwide, highlighting the urgent need for advanced therapeutic strategies. Nanoparticle-based drug delivery systems have emerged as a promising approach to improve the specificity and efficacy of anticancer treatments. This review examines three cutting-edge mesoporous silica nanoparticle (MSN)-based drug delivery to introduce novel CRC therapy, each utilizing unique functionalization strategies for targeted drug release. The first system, hyaluronidase-responsive MSN-HA/DOX, employs biotin-modified hyaluronic acid to facilitate dual-stimulus drug release in the tumor microenvironment, exhibiting enhanced in vivo tumor inhibition. The DOX/SLN-PEG-Biotin utilizes polyethylene glycol and biotin to improve drug stability and target biotin-overexpressing CRC cells, demonstrating superior anti-cancer efficacy in vitro and in vivo. Lastly, galactosylated chitosan-functionalized MSNs enable targeted delivery through asialoglycoprotein receptors, providing controlled drug release and strong therapeutic potential. Collectively, these systems highlight the advancements in nanoparticle functionalization for CRC treatment, offering a pathway to overcome the limitations of conventional chemotherapy. Further research is required to translate these innovations into clinical practice, ensuring safety and scalability.

## 1. Introduction

Colorectal cancer remains one of the most significant global health challenges, ranking as the third most diagnosed cancer and the second leading cause of cancer-related deaths worldwide. Despite advances in early detection and treatment, the five-year survival rate for CRC, especially in advanced metastatic stages, remains suboptimal. Conventional chemotherapy, though widely used, is often hindered by its non-specific nature, leading to systemic toxicity and undesirable side effects, which severely limit the therapeutic index of anticancer drugs. To address these limitations, novel drug delivery systems have been developed, with the aim of enhancing the targeted delivery of chemotherapeutic agents to the tumor site, thereby improving therapeutic outcomes and reducing off-target effects.

Nanotechnology has emerged as a promising solution in the design of targeted drug delivery, particularly for cancer therapy. Nanoparticles, due to their unique physicochemical properties such as small size, large surface area-to-volume ratio, and tunable surface chemistry, offer several advantages over traditional drug delivery methods. They can enhance drug solubility, prolong systemic circulation, and provide controlled and sustained drug release. Moreover, nanoparticles can be functionalized by targeting ligands that recognize and bind to specific receptors overexpressed on cancer cells, enabling precise drug delivery to the tumor site. Among the various types of nanoparticles investigated for cancer therapy, mesoporous silica nanoparticles have garnered considerable attention due to their highly tunable structure, biocompatibility, bioavailability and ease of surface modification.

MSNs possess a well-defined porous structure that allows for high drug loading capacity and controlled release of therapeutic agents. Their large surface area and the ability to functionalize their surface with various targeting molecules make them ideal candidates for the design of advanced drug delivery systems. Furthermore, MSNs can be engineered to respond to specific stimuli within the tumor microenvironment, such as pH, redox potential, or enzymatic activity, enabling the release of the therapeutic payload only in the presence of these triggers, thus minimizing systemic toxicity.

In this review, we discuss three recent advancements in MSN-based gated drug delivery systems designed for colorectal cancer therapy. These systems incorporate innovative functionalization strategies to enhance drug targeting, delivery, and release. The first system, hyaluronidase-responsive nanoparticles, employs biotin-modified hyaluronic acid (HA) to target cancer cells expressing hyaluronidase, a key enzyme in tumor progression. The second system utilizes dual-modified doxorubicin (DOX)-loaded silica nanoparticles with polyethylene glycol (PEG) and biotin, offering improved stability and targeted drug delivery to cancer cells with overexpressed biotin receptors. Finally, the third system explores galactosylated chitosan-functionalized MSNs, which enhance drug uptake through receptor-mediated endocytosis via asialoglycoprotein receptors, particularly targeting liver and colon cancer cells.

This report aims to provide a comprehensive overview of these advanced drug delivery systems, highlighting their design, functionality, and therapeutic potential for colorectal cancer treatment. By discussing the unique properties of each system, we aim to emphasize the potential of MSN-based targeted delivery to revolutionize the treatment of CRC, addressing the unmet need for more effective and less toxic cancer therapies. Moreover, we will discuss the future direction of this field, particularly the challenges of clinical translation and the potential for these systems to be scaled up for broad clinical application.

## 2. Hyaluronidase-Responsive Nanoparticles for Colon Cancer Therapy

*2.1 Design and Functionalization*

Zhang et al. developed a highly innovative mesoporous silica nanoparticle-based drug delivery system (MSN-HA/DOX) that utilizes the enzymatic activity of hyaluronidase (HAase) to achieve targeted drug release for colon cancer therapy. Hyaluronidase, an enzyme found to be overexpressed in several cancers, including colorectal cancer (CRC), is involved in degrading hyaluronic acid (HA), a key component of the extracellular matrix. By exploiting this enzymatic overexpression in tumor microenvironments, the MSN-HA/DOX system aims to enhance therapeutic specificity.

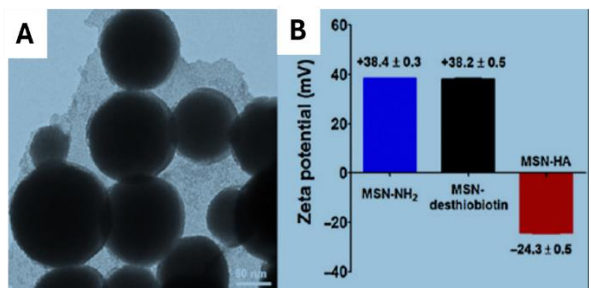

*Figure 1. (A) Transmission electron microscope images of MSN-HA; scale bar, 50 nm. (B) Zeta potentials of MSN-NH$_2$, MSN-desthiobiotin, and MSN-HA were measured. MSN-NH$_2$ and MSN-desthiobiotin showed a positive surface charge; after grafting HA, surface charge was changed to negative, indicating the successful link of HA (n=3). Reproduced with permission from [7]*

The mesoporous silica nanoparticles (MSNs) in this system were functionalized with hyaluronic acid (HA), modified with biotin for dual purposes: targeting CRC tumor cells and achieving controlled drug release. Biotin was selected due to the presence of biotin receptors that are overexpressed in various cancer cells, including colorectal cancer. This biotin-modified HA not only enables the nanoparticle to target HAase-expressing tumor cells but also acts as a 'gatekeeper' to control the release of the encapsulated drug, doxorubicin (Dox). By integrating both biotin and HA, the system achieves a synergistic effect, ensuring that drug release occurs specifically at the tumor site, driven by the enzymatic activity of HAase.

*2.2 Drug Release and Targeting*

The core mechanism of drug release in the MSN-HA/Dox system is based on a dual-stimulus approach involving hyaluronidase and biotin. The biotin-modified HA serves as a surface coating or 'gate' that blocks the release of the chemotherapeutic agent doxorubicin until the nanoparticle encounters its targeted environment, i.e., the tumor microenvironment. In this design, hyaluronidase degrades the HA coating, digesting the gatekeeping agent and releasing the encapsulated DOX. This degradation is vital because it ensures that the drug is delivered specifically at the tumor site, where hyaluronidase activity is elevated, minimizing the release of DOX in healthy tissues.

Furthermore, the biotin-functionalization contributes an additional layer of specificity, enhancing the tumor-targeting ability of the nanoparticles. Colorectal cancer cells are known to overexpress biotin receptors, which allows the biotin-modified nanoparticles to more effectively bind to the tumor cells, leading to higher drug accumulation at the cancer site. Once in proximity to the tumor cells, the HAase present in the tumor microenvironment breaks down the HA shell, triggering a targeted release of Dox directly into the cancer cells. This dual-stimulus-responsive approach not only improves drug localization but also reduces the off-target side effects typically associated with chemotherapeutic agents like doxorubicin.

*2.3 In Vitro and In Vivo Studies*

In vitro studies conducted by Zhang et al. demonstrated the effectiveness of the MSN-HA/DOX system in CRC cells. The results showed that the nanoparticles exhibited significantly enhanced drug release in the presence of hyaluronidase and biotin, which were able to selectively trigger the release of DOX. The system demonstrated high biocompatibility and efficient uptake by CRC cells, leading to enhanced therapeutic outcomes. Notably, the study showed that in the presence of HAase, the DOX release was highly controlled and precise, confirming the enzyme's role in targeting drug delivery specifically to cancer cells.

In vivo experiments further validated the potential of MSN-HA/DOX for colorectal cancer therapy. CRC xenograft models were used to test the system's therapeutic efficacy. The results were highly promising: MSN-HA/DOX showed superior anti-tumor activity compared to free DOX, with significant inhibition of tumor growth. Importantly, the targeted nature of the MSN-HA/DOX system led to minimal systemic toxicity, a common side effect in conventional chemotherapy. The dual-stimulus-responsive design, based on HAase activity and biotin targeting, ensured that the drug was delivered with high precision to the tumor site, thereby enhancing the treatment's efficacy while minimizing adverse effects on healthy tissues.

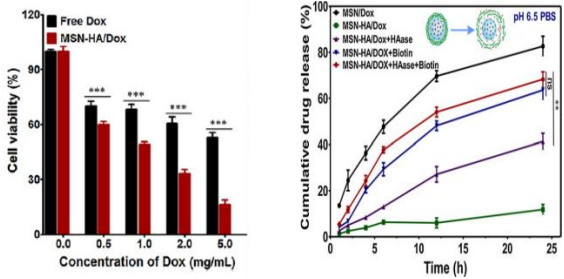

*Figure 2. The apoptotic effects of free doxorubicin and MSN-HA/DOX in Colon-26 cells were assessed by MTT assay (left). The biotin- and HAase-responsive release profiles of doxorubicin at pH 6.5 for functionalized MSNs (right). Reproduced with permission from [7]*

## 3. Doxorubicin-Loaded MSNs with Dual Modification
### 3.1 Design and Functionalization

A novel drug delivery system based on doxorubicin-loaded silica nanoparticles (DOX/SLN-PEG-Biotin) was developed by Lin et al. (2018), which integrated two key modifications to optimize the performance of the nanoparticles in colorectal cancer therapy. The first modification involved the addition of polyethylene glycol to the surface of the silica nanoparticles. PEGylation is a well-established strategy in nanomedicine, as it improves the pharmacokinetic profile of nanoparticles by increasing their stability in the circulatory system. Specifically, PEG reduces protein adsorption (opsonization), which typically leads to rapid clearance of nanoparticles from the bloodstream by the immune system. By adding PEG to the nanoparticle surface, Lin et al. effectively prolonged the circulation time of the drug-loaded nanoparticles, thereby enhancing the chances of drug accumulation at the tumor site.

A vitamin biotin was conjugated to the surface of the nanoparticles to target biotin receptors, which are overexpressed on the surface of various cancer cells, including colorectal cancer cells. Biotin-functionalized nanoparticles have shown strong targeting potential due to the increased presence of biotin receptors in tumor cells compared to normal tissues. This dual modification of the nanoparticles, combining PEG for increased circulation time and biotin for targeted delivery, aimed to maximize the therapeutic efficacy of the targeted delivery while minimizing off-target toxicity.

### 3.2 Drug Release and Cellular Uptake

The DOX/SLN-PEG-Biotin system was designed with a smart drug release mechanism, triggered by the intracellular environment of cancer cells. Once the biotin-functionalized nanoparticles reach the tumor site, they are taken up by cancer cells through receptor-mediated endocytosis, facilitated by the biotin ligands binding to the overexpressed biotin receptors on CRC cells. This ensures that the drug-loaded nanoparticles are selectively internalized by the cancer cells, reducing the chances of drug uptake by healthy cells.

Once inside the cancer cells, the nanoparticles encounter high concentrations of intracellular glutathione (GSH), which is known to be abundant in cancer cells due to their altered redox state. The DOX/SLN-PEG-Biotin system exploits this characteristic by incorporating disulfide bonds into the structure of the nanoparticles. These disulfide bonds are stable under normal physiological conditions but get cleaved in the presence of high levels of GSH. The cleavage of these bonds inside the cancer cells triggers the rapid release of doxorubicin from the nanoparticles, allowing for site-specific drug release.

This GSH-triggered release mechanism is particularly advantageous because it ensures that Dox is only released once the nanoparticles are inside the cancer cells, thereby minimizing premature drug release and reducing systemic toxicity. By using biotin for targeting and GSH for intracellular release, the DOX/SLN-PEG-Biotin system provides a highly selective and efficient means of delivering chemotherapeutic agents to CRC cells.

### 3.3 In Vitro and In Vivo Studies

In vitro studies conducted by Lin et al. demonstrated the enhanced targeting and drug release capabilities of the DOX/SLN-PEG-Biotin system. Using HCT116 colon cancer cells as the model system, the biotin-functionalized nanoparticles exhibited significantly higher cellular uptake compared to non-targeted nanoparticles. This enhanced uptake was attributed to the specific interaction between biotin ligands on the nanoparticles and biotin receptors overexpressed on the surface of CRC cells. The study confirmed that the targeted nanoparticles were efficiently internalized by the cancer cells, while the non-targeted nanoparticles showed limited cellular uptake, supporting the role of biotin in improving tumor targeting.

Furthermore, the drug release profile of the DOX/SLN-PEG-Biotin system showed accelerated release of Dox in the presence of intracellular GSH. This confirmed the GSH-triggered release mechanism, with the nanoparticles releasing their drug payload more rapidly in the reductive environment of cancer cells compared to normal cells, where GSH levels are significantly lower. This selective release mechanism further enhances the therapeutic potential of the system by ensuring that the drug is primarily released in the tumor cells, thus improving efficacy and reducing systemic side effects.

In vivo studies using a colorectal cancer mouse model further validated the therapeutic potential of the DOX/SLN-PEG-Biotin system. Mice treated with the biotin-functionalized nanoparticles showed a significant reduction in tumor growth compared to those treated with free DOX or non-targeted nanoparticles. The biotin-targeted drug delivery exhibited strong anti-cancer efficacy, with notable tumor inhibition and minimal adverse effects. Importantly, the PEGylation of the nanoparticles reduced their clearance from the bloodstream, allowing for prolonged circulation and increased drug accumulation at the tumor site.

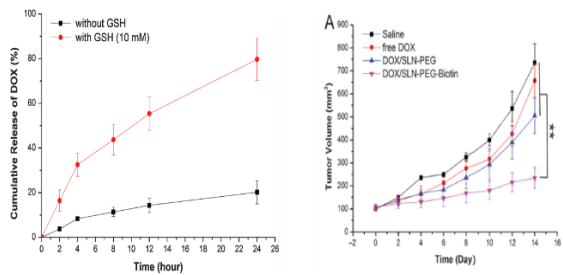

*Figure 3. In vitro drug release profile of DOX/SLN-PEG-Biotin in PBS 7.4 pH with and without 10 mM GSH (left), in vivo anti-tumor assay of DOX/SLN-PEG-Biotin with the change of tumor volume (right). Reproduced with permission from [8]*

The DOX/SLN-PEG-Biotin system's ability to combine biotin-mediated targeting, GSH-triggered drug release, and prolonged circulation time resulted in a highly effective treatment for CRC with reduced systemic toxicity. The in vivo results demonstrated that the system provided superior anti-tumor activity and minimal adverse effects compared to conventional chemotherapeutic approaches, highlighting its potential for clinical applications in cancer therapy.

## 4. Galactosylated Chitosan-functionalized MSNs for Targeted Drug Delivery

*4.1 Design and Functionalization*

Liu et al. (2018) developed an innovative drug delivery based on mesoporous silica nanoparticles functionalized with galactosylated chitosan (GC) to improve targeted drug delivery for colorectal cancer. Chitosan, a natural polysaccharide with biocompatible and biodegradable properties, was chosen as the base material for modification due to its excellent film-forming and mucoadhesive qualities. To enhance the targeting efficiency, the chitosan was modified with galactose moieties, aiming to exploit the overexpression of asialoglycoprotein receptors on CRC and liver cancer cells.

Asialoglycoprotein receptors (ASGPR) are particularly overexpressed in liver and colon cancer cells, making them an ideal target for receptor-mediated endocytosis. By attaching galactose molecules to chitosan, Liu et al. engineered the nanoparticles to have a higher affinity for these receptors. This galactose-modified chitosan coating served two critical functions: it allowed the MSNs to selectively target cancer cells via ASGPR and facilitated the internalization of the nanoparticles into CRC cells through receptor-mediated endocytosis. This design was intended to improve the specific delivery of hydrophobic chemotherapeutic agents to CRC cells, thereby enhancing therapeutic outcomes and minimizing off-target effects.

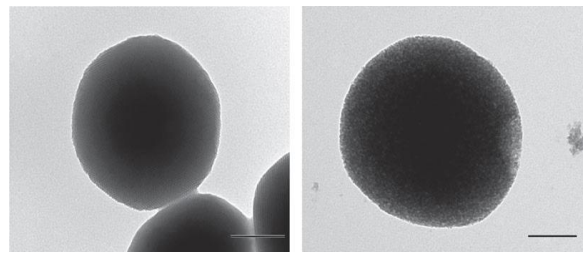

*Figure 4. TEM images of MSN (left) and MSN-NH2/GC (right). Reproduced with permission from [9]*

*4.2 Drug Release and Targeting*

The GC-functionalized MSNs were designed to provide both targeted drug delivery and controlled release of chemotherapeutic agents, optimizing the therapeutic effect. The galactose ligands on the surface of the MSNs enable the nanoparticles to bind specifically to asialoglycoprotein receptors overexpressed on CRC cells, leading to enhanced receptor-mediated endocytosis. This targeting mechanism ensures that the drug-loaded nanoparticles are preferentially taken up by CRC cells, improving the accumulation of the therapeutic agent at the tumor site.

Once inside the cancer cells, the mesoporous structure of the MSNs facilitates controlled and sustained drug release. The large surface area and tunable pore size of the MSNs provide an ideal matrix for loading hydrophobic drugs. This sustained release profile minimizes premature drug leakage during systemic circulation, ensuring that most of the drug reaches the target site before being released. By delivering the drug in a controlled manner within the tumor cells, the system enhances the therapeutic effect while minimizing systemic toxicity. This precise control over drug release helps maintain an effective concentration of the drug at the tumor site over a prolonged period, improving the efficacy of the treatment.

*4.3 In Vitro and In Vivo Studies*

Liu et al. conducted a series of in vitro studies to evaluate the effectiveness of GC-functionalized MSNs for targeted drug delivery to CRC cells. The results showed that the nanoparticles exhibited significantly enhanced cellular uptake by CRC cells, primarily through receptor-mediated endocytosis facilitated by the galactose ligands. Compared to non-functionalized MSNs, the GC-functionalized MSNs demonstrated a marked increase in internalization by cancer cells, confirming the role of ASGPR in targeting and uptake. This targeted delivery system exhibited excellent biocompatibility, with minimal cytotoxicity observed in normal cells, indicating that the functionalization with galactosylated chitosan did not compromise the safety profile of the nanoparticles.

Moreover, the GC-functionalized MSNs displayed a sustained drug release profile, further enhancing their therapeutic potential. The controlled release of the drug within CRC cells led to significant cytotoxic effects, resulting in enhanced cancer cell death compared to non-targeted systems. The in vitro studies provided compelling evidence of the system's ability to deliver chemotherapeutic agents effectively while maintaining control over drug release kinetics.

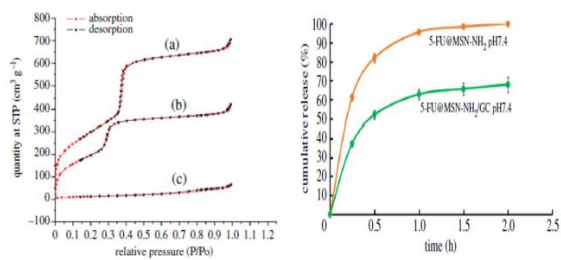

*Figure 5. Nitrogen adsorption and desorption isotherms for unmodified (a) MSN, (b) MSN-NH2 and (c) MSN-NH2/GC (left). 5-FU release profiles from 5-FU@MSN-NH2 and 5-FU@MSN-NH2/GC at pH 7.4 pH environment. Reproduced with permission from [9]*

In vivo studies further demonstrated the therapeutic efficacy of GC-functionalized MSNs in a CRC mouse model. The results showed improved drug distribution within tumor tissues, with the nanoparticles exhibiting a higher accumulation in the cancerous tissues compared to free drug formulations. The targeted delivery of the drug reduced the systemic distribution of the chemotherapeutic agent, leading to a significant decrease in off-target toxicity and adverse side effects typically associated with conventional chemotherapy.

The GC-functionalized MSNs showed superior tumor inhibition, with marked reduction in tumor size and improved survival rates in the treated mice. The in vivo experiments confirmed that the system not only enhanced drug accumulation in the tumor tissue but also provided sustained and controlled drug release, leading to better therapeutic outcomes. The reduction in systemic toxicity, coupled with the improved efficacy, highlighted the potential of this DDS for clinical applications in CRC therapy.

## 5. Conclusion

The development of advanced mesoporous silica nanoparticle-based drug delivery systems shows significant promise for enhancing colorectal cancer treatment. The reviewed systems, including hyaluronidase-responsive MSN-HA/DOX, DOX/SLN-PEG-Biotin, and galactosylated chitosan-functionalized MSNs, demonstrate how functionalization can improve drug targeting, stability, and controlled release, resulting in better therapeutic outcomes. Each approach offers distinct advantages: the MSN-HA/Dox system utilizes enzymatic activity for precise drug release in the tumor microenvironment; the DOX/SLN-PEG-Biotin system improves drug stability and targets biotin-overexpressing cancer cells; and the galactosylated chitosan-functionalized MSNs enhance drug delivery through receptor-mediated endocytosis. Collectively, these advancements address the limitations of conventional chemotherapy, such as systemic toxicity and non-specific distribution. Future research should prioritize the clinical translation of these systems, focusing on scalability, safety, and efficacy in diverse patient populations while also optimizing nanoparticle formulations and exploring combinations with other therapeutic modalities for a more integrated approach to colorectal cancer treatment.